\documentclass[aps,twocolumn,pra,showpacs,tightenlines]{revtex4-1}
\usepackage{amsmath}
\usepackage{graphicx}
\usepackage{color}
\usepackage{amsfonts}
\usepackage{txfonts}
\usepackage[colorlinks,citecolor=blue]{hyperref}

\begin{document}

\title{Quantum simulation of a three-mode optomechanical system based on the Fredkin-type interaction}
\author{Jin Liu}
\affiliation{Key Laboratory of Low-Dimensional Quantum Structures and Quantum Control of Ministry of Education, Key Laboratory for Matter Microstructure and Function of Hunan Province, Department of Physics and Synergetic Innovation Center for Quantum Effects and Applications, Hunan Normal University, Changsha 410081, China}
\author{Yue-Hui Zhou}
\affiliation{Key Laboratory of Low-Dimensional Quantum Structures and Quantum Control of Ministry of Education, Key Laboratory for Matter Microstructure and Function of Hunan Province, Department of Physics and Synergetic Innovation Center for Quantum Effects and Applications, Hunan Normal University, Changsha 410081, China}
\author{Jian Huang}
\affiliation{Key Laboratory of Low-Dimensional Quantum Structures and Quantum Control of Ministry of Education, Key Laboratory for Matter Microstructure and Function of Hunan Province, Department of Physics and Synergetic Innovation Center for Quantum Effects and Applications, Hunan Normal University, Changsha 410081, China}
\author{Jin-Feng Huang}
\affiliation{Key Laboratory of Low-Dimensional Quantum Structures and Quantum Control of Ministry of Education, Key Laboratory for Matter Microstructure and Function of Hunan Province, Department of Physics and Synergetic Innovation Center for Quantum Effects and Applications, Hunan Normal University, Changsha 410081, China}
\author{Jie-Qiao Liao}
\email{jqliao@hunnu.edu.cn}
\affiliation{Key Laboratory of Low-Dimensional Quantum Structures and Quantum Control of Ministry of Education, Key Laboratory for Matter Microstructure and Function of Hunan Province, Department of Physics and Synergetic Innovation Center for Quantum Effects and Applications, Hunan Normal University, Changsha 410081, China}
\date{\today}

\begin{abstract}
The realization of multimode optomechanical interactions in the single-photon strong-coupling regime is a desired task in cavity optomechanics, but it remains a big challenge in realistic physical systems. In this work we propose a reliable scheme to simulate a three-mode optomechanical system working in the single-photon strong-coupling regime based on the Fredkin-type interaction. This is achieved by utilizing two strong drivings to the two exchange-coupled modes in the Fredkin-type coupling involving one optical mode and two mechanical-like modes. As an application of this enhanced three-mode nonlinear optomechanical coupling, we show how to generate entangled cat states of the mechanical-like modes using the conditional displacement mechanism. The quantum coherence effects in the generated states are investigated by calculating the two-mode joint Wigner function and quantum entanglement. The influence of the system dissipation on the state generation is also considered. This work will open up a different route to the study of multimode optomechanical interactions at the single-photon level.
\end{abstract}

\maketitle

\section{Introduction}
The light-matter interaction is at the heart of the field of cavity optomechanics~\cite{Kippenberg2008rev,Aspelmeyer2012rev,Aspelmeyer2014}, and many interesting physical effects and phenomena caused by the optomechanical interactions have been demonstrated in experiments. These advances include cooling of mechanical resonators~\cite{Mancini1998PRL,Cohadon1999PRL,Kleckner2006NATURE,Corbitt2007PRL,Poggio2007PRL,Wilson2007PRL,Marquardt2007PRL,GenesVitali2008PRA,YLi2008PRB,Xia2009PRL,Tian2009PRB,Chan2011NATURE,JDTeufel2011NATURE,Clark2017NATURE}, optomechanical entanglement~\cite{Ferreira2006PRL,Vitali2007PRL,Paternostro2007PRL,Genes2008PRA,WangYD2013PRL,LTian2013PRL}, normal-mode splitting induced by strong linearized optomechanical coupling~\cite{Dobrindt2008PRL,Groblacher2009NATURE,Teufel2011NATURE,Verhagen2012NATURE}, optomechanically induced transparency~\cite{Agarwal2010PRA,Weis2010SCI,Safavi2011NATURE}, asymmetry sideband effects~\cite{Safavi2012PRL,Weinstein2014PRX,Tebbenjohanns2020PRL,LQiu2020PRL}, phonon laser~\cite{JingHui2014PRL}, and so on. Generally speaking, current studies on optomechanics focus mainly on two special cases~\cite{Aspelmeyer2014}: the strong-driving regime and the weak-driving regime. In the former case, the linearization method is used such that the linearized system can be solved exactly. In the latter case, differently, the weak-driving term is treated as a perturbation and we work in the eigen representation of the undriven optomechanical systems. In particular, in the weak-driving case, the photon number in the system is small and hence the physical effect induced by a single photon should be observable in experiments. This requirement is characterized by the single-photon strong-coupling regime, in which single-photon optomechanical effects, such as the photon blockade effect~\cite{Rabl2011PRL,Nunnenkamp2011PRL}, phonon-sideband spectrum~\cite{Liao2012PRA,Liao2013PRA,Xu2013PRA,Hong2013PRA}, and generation of cat states~\cite{Marshall2003PRL,Liao2016PRL}, can be realized in experiments. Though much effort has been devoted to the enhancement of the single-photon optomechanical effects~\cite{AXuerebPRL2012,AJRimberg2014NJP,TTHeikkila2014PRL,JMPirkkalainen2015NC,JQLiao2014NJP,JQLiao2015PRA,XYLue2015PRL,Lemonde2016NC,ZWang2017NC}, the observation of single-photon optomechanical effects has remained a big challenge.

In parallel with the extensive studies of optomechanical couplings in the single-photon strong-coupling regime, considerable advances have also been made in multimode optomechanics~\cite{Bhattacharya2008PRA,Nair2016PRA,Spethmann2016NP,Massel2017PRA,Xu2017PRL,GilSantos2017PRL,Nielsen2016PNAS}. This is because multimode optomechanical systems provide a promising platform to study macroscopic quantum coherence involving multiple mechanical modes~\cite{XuXW2013PRA,LiaoJQ2014PRA,WangM2016PRA,OckeloenKorppi2016NATURE}. For example, both theoretical and experimental advances have been made in the generation of entangled states involving multiple mechanical resonators~\cite{Akram2013NJP}. In addition, quantum synchronization of mechanical motion of multiple resonators has also been studied~\cite{Mari2013PRL,Matheny2014PRL,WLLin2020PRA}. In particular, recent studies have shown that simultaneous ground-state cooling of multiple mechanical mode can be realized~\cite{Genes2008NJP,DGLai2018PRA,DGLai2019PRA,Sommer2019PRL,Ockeloen2019PRA}. These advances open a new realm to multimode quantum optomechanics, which will have wide applications in both the study of the fundamentals of quantum mechanics and modern quantum technologies.

Motivated by the recent research interest in both the single-photon strong-coupling regime and multimode optomechanics, it is desiable to study the multimode optomechanical interactions in the single-photon strong-coupling regime. Currently, the single-photon strong-coupling or ultrastrong-coupling regime of the optomechanical interaction has not been experimentally realized~\cite{Rabl2011PRL,Nunnenkamp2011PRL,Liao2012PRA,Liao2013PRA,DHu2015PRA,Garziano2015PRA,VMacr2016PRA,JQLiao2016PRA}. In this situation, quantum simulation~\cite{Buluta2009Science,Georgescu2014RMP} might be a powerful way to explore the optomechanical interactions in the single-photon strong-coupling regime, because one of the motivations of quantum simulation is to simulate the experimentally inaccessible physical effects with other experimentally accessible systems.

In this paper we propose a reliable scheme to implement the quantum simulation of a three-mode optomechanical model based on the Fredkin-type interaction~\cite{Milburn1989PRL,Patel2016SA,GAO2019NATURE} which involves one optical mode and two mechanical-like modes. By introducing strong drivings to the two mechanical-like modes, the Fredkin interaction will lead to a three-mode optomechanical interaction with two enhanced coupling strengths. Here the three bosonic modes play the role of an optical mode (the conditional controller mode in the Fredkin interaction) and two mechanical modes (the two modes involving the exchange coupling). In particular, the simulated three-mode optomechanical interaction can enter the single-photon strong-coupling regime and even the ultrastrong-coupling regime. As an application of the enhanced three-mode optomechanical interaction, we study the generation of entangled cat states in the two mechanical-like modes. The coherence effects of the generated states are investigated by calculating the joint Wigner function and quantum entanglement between the two mechanical-like modes.

The rest of this work is organized as follows. In Sec.~\ref{model} we present the physical model and the Hamiltonians. We also derive an effective three-mode optomechanical Hamiltonian. In Sec.~\ref{stategeneration} we study the generation of entangled cat states with the approximate Hamiltonian and investigate the nonclassical properties of the generated states. In addition, we verify the validity of the approximate Hamiltonian. In Sec.~\ref{opensystem} we study the influence of the system dissipations on the state generation. In Sec.~\ref{Discussion} we present a discussion of the experimental implementation of our scheme. We give a brief summary in Sec.~\ref{Conclusion}.

\section{MODEL AND SIMULATED OPTOMECHANICAL HAMILTONIAN \label{model}}
\begin{figure}[tbp]
\centering
\includegraphics[scale=0.52]{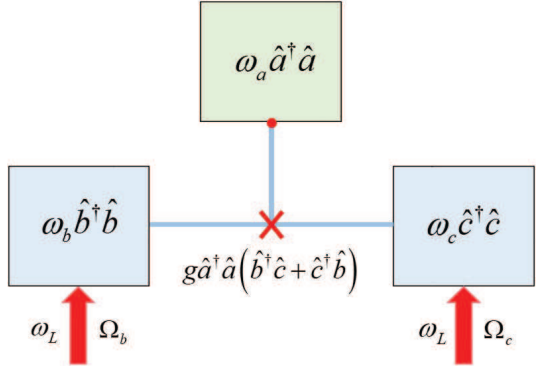}
\caption{Schematic of the Fredkin-type interacting system formed by a bosonic mode $a$ (with a resonance frequency $\omega_{a}$) coupled to two bosonic modes $b$ and $c$ (with the resonance frequencies $\omega_{b}$ and $\omega_{c}$). Modes $b$ and $c$ are driven by two strong laser fields with the same frequency $\omega_{L}$ and individual driving amplitudes $\Omega_{b}$ and $\Omega_{c}$, respectively.}
\label{modelpicture}
\end{figure}

We consider a Fredkin-type interacting system (Fig.~\ref{modelpicture}), which is composed of three bosonic modes described by the annihilation (creation) operators $\hat{a}$ ($\hat{a}^{\dag}$), $\hat{b}$ ($\hat{b}^{\dag}$), and $\hat{c}$ ($\hat{c}^{\dag}$), with the corresponding resonance frequencies $\omega_{a}$, $\omega _{b}$, and $\omega _{c}$. Here, two (e.g., $b$ and $c$) of the three modes are coupled with each other through a beam-splitter-type interaction, where the coupling strength is proportional to the excitation number in the third mode $(a)$. The Fredkin-type interaction Hamiltonian takes the form $ \hat{a}^{\dag }\hat{a}(\hat{b}^{\dag }\hat{c}+\hat{c}^{\dag}\hat{b})$. This interaction has been suggested to implement a quantum logic gate, and the experimental implementation of the Fredkin-type interaction has been suggested in optical systems~\cite{Milburn1989PRL} and realized in both an optical system~\cite{Patel2016SA} and a circuit-QED system~\cite{GAO2019NATURE}. In our scheme, modes $b$ and $c$ are strongly driven by two laser fields with the same driving frequency $\omega_{L}$ and individual driving amplitudes $\Omega_{b}$ and $\Omega_{c}$, respectively. For simulation of a three-mode optomechanical interaction, here mode $a$ plays the role of an optical mode, while modes $b$ and $c$ play the role of two mechanical modes. The Hamiltonian of the system reads (with $\hbar=1$)
\begin{eqnarray}
\hat{H}(t)&=&\omega_{a}\hat{a}^{\dag}\hat{a}+\omega_{b}\hat{b}^{\dag }\hat{b}+\omega_{c}\hat{c}^{\dag }\hat{c}+g\hat{a}^{\dag }\hat{a}(\hat{b}^{\dag }\hat{c}+\hat{c}^{\dag}\hat{b})\notag\\
&&+\Omega_{b}(\hat{b}^{\dag}e^{-i\omega_{L}t}+\hat{b}e^{i\omega_{L}t})+\Omega_{c}(\hat{c}^{\dag}e^{-i\omega_{L}t}+\hat{c}e^{i\omega_{L}t}),\label{mainHamiltonian}
\end{eqnarray}
where $g$ is the coupling strength of the Fredkin-type interaction. In a rotating frame with respect to $\hat{H}_{0}=\omega_{L}(\hat{b}^{\dag}\hat{b}+\hat{c}^{\dag}\hat{c})$, the Hamiltonian~(\ref{mainHamiltonian}) becomes
\begin{eqnarray}
\hat{H}_{I}&=&\omega_{a}\hat{a}^{\dag}\hat{a}+\Delta_{b}\hat{b}^{\dag}\hat{b}+\Delta_{c}\hat{c}^{\dag}\hat{c}+g\hat{a}^{\dag}\hat{a}(\hat{b}^{\dag}\hat{c}+\hat{c}^{\dag}\hat{b})\notag\\
&&+\Omega_{b}(\hat{b}^{\dag}+\hat{b})+\Omega_{c}(\hat{c}^{\dag}+\hat{c}),\label{rotatedHamiltonian}
\end{eqnarray}
where $\Delta_{b}=\omega_{b}-\omega_{L}$ and $\Delta_{c}=\omega_{c}-\omega_{L}$ are the driving detunings of modes $b$ and $c$, respectively.

In the strong-driving case, the average excitation numbers in modes $b$ and $c$ are large and hence we can express the operators of modes $b$ and $c$ as the sum of their average values and quantum fluctuations. This procedure can be conducted by performing the displacement transformation on modes $b$ and $c$. To this end, we introduce the displacement operators $\hat{D}_{b}(\xi_{b})=\exp(\xi_{b}\hat{b}^{\dag}-\xi_{b}^{\ast}\hat{b})$ and $\hat{D}_{c}(\xi_{c})=\exp(\xi_{c}\hat{c}^{\dag}-\xi_{c}^{\ast}\hat{c})$, where $\xi_{b}$ and $\xi_{c}$ are the displacement amplitudes of modes $b$ and $c$, respectively. By choosing the displacement amplitudes $\xi_{b}=-\Omega_{b}/\Delta_{b}$ and $\xi_{c}=-\Omega_{c}/\Delta_{c}$, we perform the displacement transformation on Hamiltonian~(\ref{rotatedHamiltonian}) and obtain the transformed Hamiltonian as
\begin{eqnarray}
\hat{H}_{\text{ext}}&=&\hat{D}_{c}^{\dag}(\xi_{c})\hat{D}_{b}^{\dag}(\xi_{b})\hat{H}_{I}\hat{D}_{b}(\xi_{b})\hat{D}_{c}(\xi_{c})\notag\\
&=&\tilde{\omega}_{a}\hat{a}^{\dag}\hat{a}+\Delta_{b}\hat{b}^{\dag}\hat{b}+\Delta_{c}\hat{c}^{\dag}\hat{c}+g\hat{a}^{\dag}\hat{a}(\hat{b}^{\dag}\hat{c}+\hat{c}^{\dag}\hat{b})\notag\\
&&+g_{b}\hat{a}^{\dag}\hat{a}(\hat{b}^{\dag}+\hat{b})+g_{c}\hat{a}^{\dag}\hat{a}(\hat{c}^{\dag }+\hat{c}),\label{ExactHamiltonian}
\end{eqnarray}
where we introduce the normalized frequency $\tilde{\omega}_{a}=\omega_{a}+2g\xi_{b}\xi_{c}$ of mode $a$, and the modulated coupling strengths $g_{b}=g\xi_{c} $ and $g_{c}=g\xi_{b} $.

Our task in this work is to simulate a three-mode optomechanical interaction including one optical mode and two mechanical-like modes based on the present physical system. We can find from Eq.~(\ref{ExactHamiltonian}) that if the three-mode coupling term  $g\hat{a}^{\dag}\hat{a}(\hat{b}^{\dag}\hat{c}+\hat{c}^{\dag }\hat{b})$ in the  Hamiltonian~(\ref{ExactHamiltonian}) can be approximately ignored under proper parameter conditions, then we get a standard three-mode optomechanical-interaction Hamiltonian. To analyze the parameter conditions related to the approximation more clearly, we work in the interaction picture with respect to $\hat{H}_{\text{ext}}^{0}=\tilde{\omega}_{a}\hat{a}^{\dag}\hat{a}+\Delta_{b}\hat{b}^{\dag}\hat{b}+\Delta_{c}\hat{c}^{\dag}\hat{c}$. Then the Hamiltonian~(\ref{ExactHamiltonian}) becomes
\begin{eqnarray}
\hat{H}^{I}_{\text{ext}}&=&g\hat{a}^{\dag}\hat{a}\hat{b}^{\dag}\hat{c} e^{i(\Delta_{b}-\Delta_{c})t}+g\hat{a}^{\dag}\hat{a}\hat{c}^{\dag}\hat{b}e^{-i(\Delta_{b}-\Delta_{c})t} \notag\\
&&+g_{b}\hat{a}^{\dag}\hat{a}\hat{b}^{\dag}e^{i\Delta_{b}t}+g_{b}\hat{a}^{\dag}\hat{a}\hat{b}e^{-i\Delta_{b}t} \notag\\
&&+g_{c}\hat{a}^{\dag}\hat{a}\hat{c}^{\dag }e^{i\Delta_{c}t}+g_{c}\hat{a}^{\dag}\hat{a}\hat{c}e^{-i\Delta_{c}t}. \label{InteractionExactHamiltonian}
\end{eqnarray}
In Eq.~(\ref{InteractionExactHamiltonian}), we can compare the contribution of the terms relating to $g\hat{a}^{\dag}\hat{a}\hat{b}^{\dag}\hat{c} e^{i(\Delta_{b}-\Delta_{c})t}$, $g_{b}\hat{a}^{\dag}\hat{a}\hat{b}^{\dag}e^{i\Delta_{b}t}$, and $g_{c}\hat{a}^{\dag}\hat{a}\hat{c}^{\dag}e^{i\Delta_{c}t}$. We find that, under the conditions
\begin{eqnarray}
\bigg| g_{b}\frac{ n_{a} \sqrt{n_{b}}}{\Delta_{b}} \bigg| \gg \bigg| \frac{g n_{a} \sqrt{n_{b}}}{\Delta_{b}-\Delta_{c}} \sqrt{n_c} \bigg|, \notag \\
\bigg| g_{c}\frac{ n_{a} \sqrt{n_{c}}}{\Delta_{c}}\bigg|\gg\bigg|\frac{g n_{a} \sqrt{n_{c}}}{\Delta_{b}-\Delta_{c}} \sqrt{n_b}\bigg|, \label{ApproximationCondition}
\end{eqnarray}
the terms in the first of Eqs.~(\ref{InteractionExactHamiltonian}) can be approximately discarded.
Here, $n_{a}$, $n_{b}$, and $n_{c}$ are the maximally contributed excitation numbers in modes $a$, $b$, and $c$, respectively. Based on the relations $g_{b}=g\xi_{c}$ and $g_{c}=g\xi_{b}$, we can simplify the approximation conditions as
\begin{eqnarray}
|\xi_{c}|= \bigg|\frac{\Omega_{c}}{\Delta_{c}}\bigg|\gg \frac{\sqrt{n_{c}}}{|1-\Delta_{c}/\Delta_{b}|} , 
|\xi_{b}|= \bigg| \frac{\Omega_{b}}{\Delta_{b}} \bigg| \gg\frac{\sqrt{n_{b}}}{|\Delta_{b}/\Delta_{c}-1|}.\label{condparaappro}
\end{eqnarray}
Here we can find that the approximation conditions are related to the ratio $\Delta_{b}/\Delta_{c}$. In Sec.~\ref{idilitysec}, we will analyze the dependence of the approximation quality on the parameter conditions in detail.

Now let us return to the displacement representation. We consider the few-excitation case and assume that the parameter conditions in Eq.~(\ref{condparaappro}) are satisfied. Then the three-mode coupling term $g\hat{a}^{\dag}\hat{a}(\hat{b}^{\dag}\hat{c}+\hat{c}^{\dag }\hat{b})$ can be ignored, and we obtain the approximate Hamiltonian as
\begin{eqnarray}
\hat{H}_{\text{app}}&\approx&\tilde{\omega}_{a}\hat{a}^{\dag }\hat{a}+\Delta _{b}\hat{b}^{\dag}\hat{b}+\Delta _{c}\hat{c}^{\dag }\hat{c}\notag \\
&&+g_{b}\hat{a}^{\dag}\hat{a}(\hat{b}^{\dag }+\hat{b})+g_{c}\hat{a}^{\dag}\hat{a}(\hat{c}^{\dag}+\hat{c}).\label{EffectiveHamiltonian}
\end{eqnarray}
The Hamiltonian $\hat{H}_{\text{app}}$ takes the standard form of the three-mode optomechanical interaction with one optical mode $a$ and two mechanical-like modes $b$ and $c$, where $\Delta_{b}$ and $\Delta_{c}$ are effective mechanical frequencies and $g_{b}$ and $g_{c}$ are single-photon optomechanical-coupling strengths. Here, the two coupling strengths can be largely enhanced by choosing large displacement amplitudes $\xi_{b}$ and $\xi_{c}$ in the strong-driving cases. Then, this three-mode optomechanical model can enter the single-photon strong-coupling and even ultrastrong-coupling regimes. Note that the typical ultrastrong optomechanics can be implemented by driving one of the two mechanical-like modes~\cite{XianLiYin2020}.

\section{Generation of entangled cat states \label{stategeneration}}

As an application of the enhanced three-mode optomechanical interaction, we study how to generate entangled cat states based on the dynamical evolution of the system. We also investigate quantum effects in the generated states by analyzing the joint Wigner function and the degree of entanglement. In addition, the validity of the approximate Hamiltonian~(\ref{EffectiveHamiltonian}) is evaluated.

\subsection{State generation under the approximate Hamiltonian}

To calculate the expression of the generated state, we first diagonalize the approximate Hamiltonian~(\ref{EffectiveHamiltonian}). To this end, we introduce the unitary displacement operators $\hat{D}_{b}(\hat{\eta}_{b})=\exp[\hat{\eta}_{b}(\hat{b}^{\dag }-\hat{b})]$ and $\hat{D}_{c}(\hat{\eta}_{c})=\exp[\hat{\eta}_{c}(\hat{c}^{\dag }-\hat{c})]$, where $\hat{\eta}_{b}$ and $\hat{\eta}_{c}$ are functions of $\hat{a}^{\dag }\hat{a}$,
\begin{subequations}
\begin{align}
\hat{\eta}_{b}(\hat{a}^{\dag}\hat{a})=\frac{g\Omega_{c}}{\Delta_{b}\Delta_{c}}\hat{a}^{\dag }\hat{a}=\sum_{n=0}^{\infty}\eta_{b}(n)|n\rangle_{a a}\langle n|,\\
\hat{\eta}_{c}(\hat{a}^{\dag}\hat{a})=\frac{g\Omega_{b}}{\Delta_{b}\Delta_{c}}\hat{a}^{\dag }\hat{a}=\sum_{n=0}^{\infty}\eta_{c}(n)|n\rangle_{a a}\langle n|,
\end{align}
\end{subequations}
with $\eta_{b}(n)=gn\Omega_{c}/\Delta_{b}\Delta_{c}$ and $\eta_{c}(n)=gn\Omega_{b}/\Delta_{b}\Delta_{c}$ for natural number $n$. By performing the displacement transformation on Eq.~(\ref{EffectiveHamiltonian}), we obtain the diagonalized Hamiltonian as
\begin{eqnarray}
\hat{\tilde{H}}_{\text{app}}&=&\hat{D}_{c}^{\dag}(\hat{\eta}_{c})\hat{D}_{b}^{\dag}(\hat{\eta}_{b})\hat{H}_{\text{app}}\hat{D}_{b}(\hat{\eta}_{b})\hat{D}_{c}(\hat{\eta}_{c})\notag \\
&=&\tilde{\omega}_{a}\hat{a}^{\dag}\hat{a}+\Delta_{b}\hat{b}^{\dag }\hat{b}+\Delta _{c}\hat{c}^{\dag }\hat{c}-(\Delta_{b}\hat{\eta}_{b}^{2}+\Delta_{c}\hat{\eta}_{c}^{2}).\label{DiagonizedEffectiveHamiltonian}
\end{eqnarray}
Assume that the system is initially in the state
\begin{equation}
\left\vert\psi(0)\right\rangle=\frac{1}{\sqrt{2}}(\left\vert0\right\rangle_{a}+\left\vert1\right\rangle_{a})|0\rangle_{b}|0\rangle_{c},\label{InitialState}
\end{equation}
where $\left\vert0\right\rangle_{a}$ and $\left\vert1\right\rangle_{a}$ denote the vacuum state and single-excitation state of mode $a$, respectively, while $|0\rangle_{b}$ ($|0\rangle_{c}$) is the vacuum state of mode $b$ ($c$).
The state of the system at time $t$ can be obtained according to the relation
\begin{eqnarray}
\left\vert\psi(t)\right\rangle &=&e^{-i\hat{H}_\text{app}t}\left\vert \psi(0)\right\rangle \notag \\
&=&\hat{D}_{b}(\hat{\eta}_{b})\hat{D}_{c}(\hat{\eta}_{c})e^{-i\hat{\tilde{H}}_{\text{app}}t}\hat{D}_{c}^{\dag}(\hat{\eta}_{c})\hat{D}_{b}^{\dag}(\hat{\eta}_{b})\left\vert \psi(0)\right\rangle.
\end{eqnarray}
After a detailed calculation, we obtain
\begin{eqnarray}
\left\vert\psi(t)\right\rangle  &=&\frac{1}{\sqrt{2}}\left(\left\vert 0\right\rangle _{a}\left\vert 0\right\rangle_{b}\left\vert 0\right\rangle_{c}+e^{-i\theta_{1}(t)}\left\vert 1\right\rangle _{a}\left\vert \alpha_{1}(t)\right\rangle _{b}\left\vert \beta_{1}(t)\right\rangle _{c}\right),\label{psi(t)} \nonumber\\
\end{eqnarray}
where we introduce the phase
\begin{equation}
\theta_{1}(t)=\tilde{\omega}_{a}t+\Lambda_{1}t+\phi_{1}(t),
\end{equation}
with
\begin{subequations}
\begin{align}
\Lambda_{1}=&\Delta_{b}\eta_{b}^{2}(1)+\Delta_{c}\eta_{c}^{2}(1)+2g[\xi_{c}\eta_{b}(1)+\xi_{b}\eta_{c}(1)],\\
\phi_{1}(t)=&\eta_{b}^{2}(1)\text{sin}(\Delta_{b}t)+\eta_{c}^{2}(1)\text{sin}(\Delta_{c}t).
\end{align}
\end{subequations}
The coherent-state amplitudes in Eq.~(\ref{psi(t)}) are given by
\begin{subequations}
\begin{align}
\alpha_{1}(t)=&\eta_{b}(1)(1-e^{-i\Delta_{b}t}),\\
\beta_{1}(t)=&\eta_{c}(1)(1-e^{-i\Delta_{c}t}),
\end{align}
\end{subequations}
where $\eta_{b}(1)=g\Omega_{c}/\Delta_{b}\Delta_{c}$ and $\eta_{c}(1)=g\Omega_{b}/\Delta_{b}\Delta_{c}$.

For generation of two-mode entangled cat states, we express the states of mode $a$ with the bases $|\pm\rangle=(\left\vert0\right\rangle_{a}\pm\left\vert1\right\rangle_{a}) /\sqrt{2}$, then the generated state $\left\vert\psi(t)\right\rangle$ becomes
\begin{equation}
|\psi(t)\rangle=\frac{1}{2\mathcal{M}_{+}(t)}|+\rangle\left|\psi_{+}(t)\right\rangle+\frac{1}{2\mathcal{M}_{-}(t)}|-\rangle\left|\psi_{-}(t)\right\rangle,\label{psi(t)1}
\end{equation}
where we introduce the entangled cat states for modes $b$ and $c$ as
\begin{equation}
\left\vert\psi_{\pm}(t)\right\rangle=\mathcal{M}_{\pm}(t)\left( \left\vert 0 \right\rangle_{b}\left\vert 0 \right\rangle_{c}\pm e^{-i\theta_{1}(t)}\left\vert \alpha_{1}(t)\right\rangle_{b}\left\vert \beta_{1}(t)\right\rangle_{c}\right),\label{EffectiveEntangledCoherentStates}
\end{equation}
with the normalization constants
\begin{equation}
\mathcal{M}_{\pm}(t)=\left(2 \{1 \pm e^{-[|\alpha_{1}(t)|^{2}+|\beta_{1}(t)|^{2}]/2}\text{cos}[\theta_{1}(t)] \}\right)^{-1/2}.\label{NormalizationConstantsM}
\end{equation}

By performing a measurement of mode $a$ in the basis states $\left\vert\pm \right\rangle$, modes $b$ and $c$ will collapse into the entangled cat states $\left\vert\psi_{\pm }(t)\right\rangle.$
The corresponding probabilities for the detection are
\begin{equation}
\mathcal{P}_{\pm}(t)=\frac{1}{4\left\vert \mathcal{M}_{\pm}(t)\right\vert ^{2}},\label{EffectiveProbabilities}
\end{equation}
which are the success probabilities for generation of entangled cat states $|\psi_{\pm}(t)\rangle$, respectively.

\subsection{Logarithmic negativity of the entangled cat states}
\begin{figure}[tbp]
\centering
\includegraphics[scale=0.58]{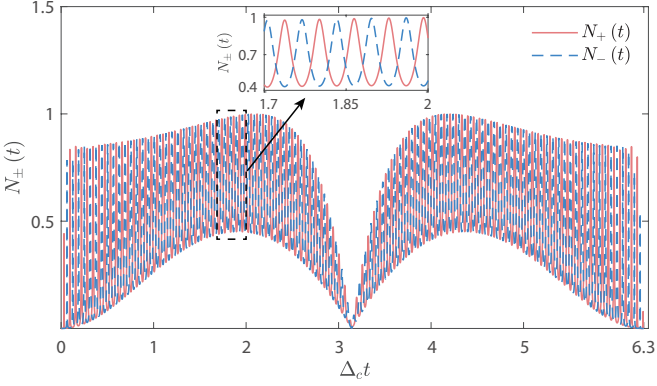}
\caption{Dynamics of the logarithmic negativities $N_{\pm}(t)$ for states $\left|\psi_{\pm}(t)\right\rangle$ in Eq.~(\ref{EffectiveEntangledCoherentStates}). The inset shows a close-up plot of $N_{\pm}(t)$ in the middle duration of the half-period. The other parameters are $\omega_{a}/\Delta_{c}=0.1$, $\Omega_{b}/\Delta_{c}=\Omega_{c}/\Delta_{c}=100$, $g/\Delta_{c}=0.01$, and $\Delta_{b}/\Delta_{c}=2$.}
\label{Entanglement(closed)}
\end{figure}

The degree of entanglement of the generated entangled cat states $|\psi_{\pm}(t)\rangle$ can be quantized by calculating the logarithmic negativity~\cite{Vidal2002,Plenio2005}. For a two-partite system described by the density matrix $\hat{\rho}$, the logarithmic negativity is
defined by
\begin{equation}
N=\log_{2}\left\vert\left\vert \hat{\rho}^{T_{c}}\right\vert\right\vert_{1},\label{LogarithmicNegativity}
\end{equation}
where $T_{c}$ denotes the partial transpose of the density matrix $\hat{\rho}$ with respect to mode $c$, and the trace norm $\left\vert \left\vert \hat{\rho}^{T_{c}}\right\vert \right\vert_{1} $ is defined by
\begin{equation}
\left\vert\left\vert \hat{\rho}^{T_{c}}\right\vert\right\vert_{1}=\text{Tr}\left[\sqrt{(\hat{\rho}^{T_{c}})^{\dagger}\hat{\rho}^{T_{c}}}\right].\label{TraceNorm}
\end{equation}

Below, we expand the generated states $|\psi_{\pm}(t)\rangle$ in the Fock space, and then calculate their logarithmic negativity. In Fig.~\ref{Entanglement(closed)} we show the dynamics of the logarithmic negativities $N_{\pm}(t)$ for the generated states $\left|\psi_{\pm}(t)\right\rangle$. We find that the logarithmic negativities $N_{\pm}(t)$ exhibit fast oscillations in a whole period; these oscillations are caused by the high-frequency term in the phase factor $\theta_{1}(t)$. For the parameters $\Delta_{b}=2\Delta_{c}$, we find either $|\alpha_{1}(t)|=0$ or $|\beta_{1}(t)|=0$ at $\Delta_{c}t=n\pi$; then modes $b$ and $c$ decouple from each other so that there is no entanglement, i.e., the logarithmic negativity is 0. In the middle of a half-period, the amplitude of the oscillation is reduced, and the logarithmic negativities $N_{\pm}(t)$ reach the maximum values (see the inset in Fig.~\ref{Entanglement(closed)}). We can also see that the peak positions of $N_{\pm}(t)$ appear alternately.

\begin{figure*}[tbp]
\centering
\includegraphics[scale=0.50]{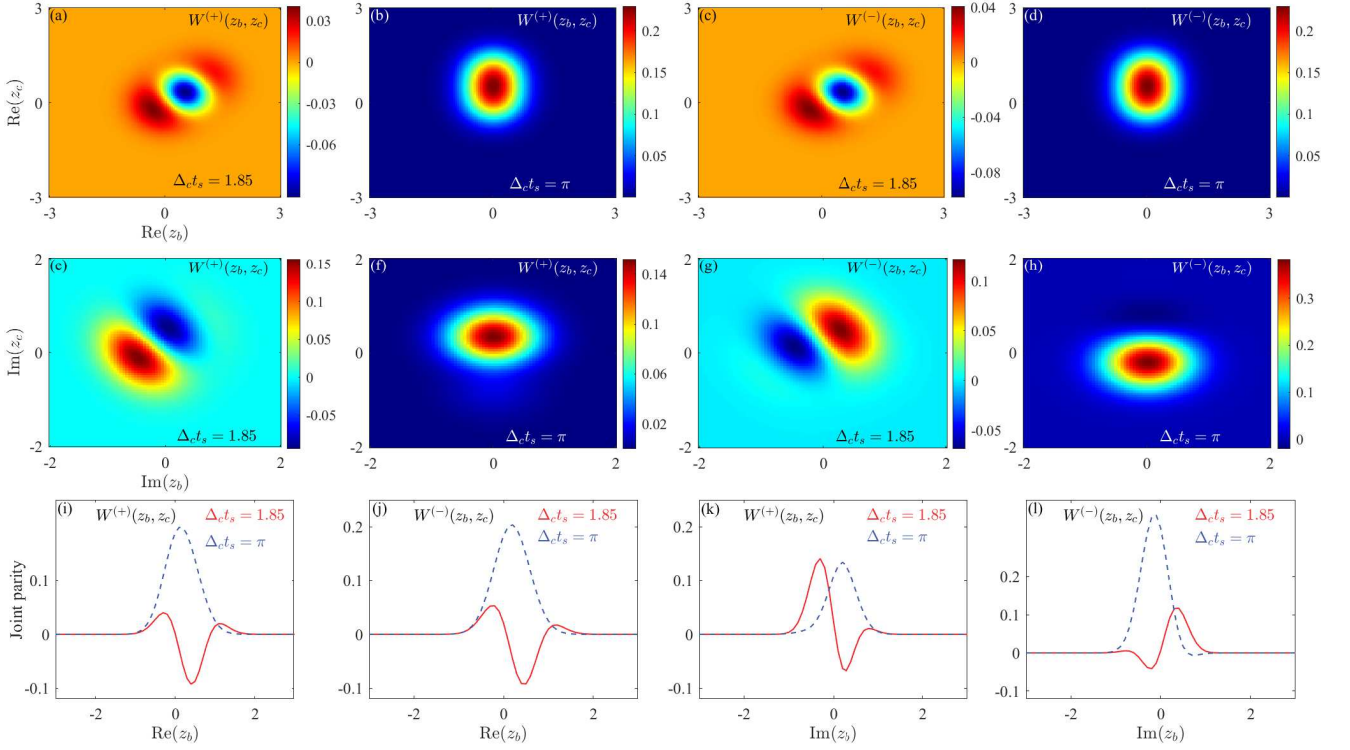}
\caption{Two-mode joint Wigner function $W^{(+)}(z_{b},z_{c})$ $[W^{(-)}(z_{b},z_{c})]$ along either the $\text{Re}(z_{b})-\text{Re}(z_{c})$ plane at times (a) [(c)] $\Delta_{c}t_{s}=1.85$ and (b) [(d)] $\Delta_{c}t_{s}=\pi$ or the $\text{Im}(z_{b})-\text{Im}(z_{c})$ plane at times (e) [(g)] $\Delta_{c}t_{s}=1.85$ and (f) [(h)] $\Delta_{c}t_{s}=\pi$. For the entangled cat state $\left\vert\alpha_{1}(\Delta_{c}t_{s})\right\vert\approx0.96$ and $\left\vert\beta_{1}(\Delta_{c}t_{s})\right\vert\approx0.79$ at $\Delta_{c}t_{s}=1.85$. (i) and (j) Diagonal line-cuts of the joint Wigner functions $W^{(\pm)}(z_{b},z_{c})$ from (a)-(d) along $\text{Re}(z_{b})=\text{Re}(z_{c})$ with $\text{Im}(z_{b})=0.5$ and $\text{Im}(z_{c})=0.4$ at different times ($\Delta_{c}t_{s}=1.85$ for the solid line and $\Delta_{c}t_{s}=\pi$ for the dashed line). (k) and (l) Diagonal line cuts of the joint Wigner functions $W^{(\pm)}(z_{b},z_{c})$ from (e)-(h) along $\text{Im}(z_{b})=\text{Im}(z_{c})$ with $\text{Re}(z_{b})=0.5$ and $\text{Re}(z_{c})=0$ at different times ($\Delta_{c}t_{s}=1.85$ for the solid line and $\Delta_{c}t_{s}=\pi$ for the dashed line). The other parameters are $\omega_{a}/\Delta_{c}=0.1$, $\Omega_{b}/\Delta_{c}=\Omega_{c}/\Delta_{c}=100$, $g/\Delta_{c}=0.01$, and $\Delta_{b}/\Delta_{c}=2$.}
\label{Wignerfunction(closed)}
\end{figure*}

\subsection{Joint Wigner function of the entangled cat states}

In order to show the quantum interference and coherence effects in the generated entangled cat states, we calculate the joint Wigner function of modes $b$ and $c$. The measurement of the joint Wigner function gives the quantum state tomography of the two-mode system, which shows the details of quantum properties in the two-mode system. For a two-bosonic-mode ($b$ and $c$) system, the joint Wigner function is defined by~\cite{CWang2016,Zhong2016,HuangJian2020}
\begin{eqnarray}
&&W(z_{b},z_{c})=\frac{4}{\pi^{2}}\left\langle\hat{D}_{b}(z_{b})(-1)^{\hat{b}^{\dag}\hat{b}}\hat{D}_{b}^{\dag}(z_{b})\hat{D}_{c}(z_{c})(-1)^{\hat{c}^{\dag}\hat{c}}\hat{D}_{c}^{\dag }(z_{c})\right\rangle,\notag \\&&
\end{eqnarray}
where $\hat{D}_{b}(z_{b})=\exp({z_{b}\hat{b}^{\dag}-z_{b}^{\ast}\hat{b}})$ and $\hat{D}_{c}(z_{c})=\exp({z_{c}\hat{c}^{\dag}-z_{c}^{\ast}\hat{c}})$ are the displacement operators of the two bosonic modes $b$ and $c$, respectively, with $z_{b} $ and $z_{c}$ being the complex parameters defining the coordinates in the joint phase space. Here $W(z_{b},z_{c})$ is a function in the four-dimensional (4D) phase space [$\text{Re}(z_{b}),\text{Im}(z_{b}),\text{Re}(z_{c})$, and $\text{Im}(z_{c})$]. The value of $W(z_{b},z_{c})$ can be measured from the expectation value of the joint parity operator $(-1)^{\hat{b}^{\dag }\hat{b}+\hat{c}^{\dag }\hat{c}}$ after independent displacements $\hat{D}_{b}(z_{b})$ and $\hat{D}_{c}(z_{c})$. Therefore, we will use the displaced joint parity function to the demonstrate nonclassical property between the two bosonic modes $b$ and $c$.

To explore the feature in the Wigner functions corresponding to the states at special times in Fig~\ref{Entanglement(closed)}, we plot the 4D Wigner functions of the states $\left\vert\psi_{\pm}(t)\right\rangle$ and display their two-dimensional cuts along the $\text{Re}(z_{b})$-$\text{Re}(z_{c})$ plane and the  $\text{Im}(z_{b})$-$\text{Im}(z_{c})$ plane at different certain times, as shown in Fig.~\ref{Wignerfunction(closed)}. In Figs~\ref{Wignerfunction(closed)}(a),~\ref{Wignerfunction(closed)}(c),~\ref{Wignerfunction(closed)}(e), and~\ref{Wignerfunction(closed)}(g) we show the Wigner functions for the states at time $\Delta_{c}t_{s}\approx1.85$. The entanglement is close to the maximal value at this moment. Here we can see that the Wigner functions contain positive and negative Gaussian balls in Figs~\ref{Wignerfunction(closed)}(a) and~\ref{Wignerfunction(closed)}(c) and some stripes with positive and negative values in Figs~\ref{Wignerfunction(closed)}(e) and~\ref{Wignerfunction(closed)}(g). These characteristics show the quantum properties of entangled coherent states. In panels~\ref{Wignerfunction(closed)}(b),~\ref{Wignerfunction(closed)}(d),~\ref{Wignerfunction(closed)}(f), and~\ref{Wignerfunction(closed)}(h) we show the Wigner functions for the states at the decoupling time $\Delta_{c}t_{s}=\pi$. At this moment, there is no entanglement between the two modes $b$ and $c$. Figures~\ref{Wignerfunction(closed)}(b),~\ref{Wignerfunction(closed)}(d),~\ref{Wignerfunction(closed)}(f), and~\ref{Wignerfunction(closed)}(h) show that all of the above interesting characteristics have disappeared. To show the details of these interesting characteristics, we plot in Figs.~\ref{Wignerfunction(closed)}(i)-(l) the diagonal line cuts of these Wigner functions. We can see that the diagonal line cuts exhibit some oscillation for the entangled cat states at $\Delta_{c}t_{s}\approx1.85$. For the disentangled state at $\Delta_{c}t_{s}=\pi$, the diagonal line-cuts only show a peak distribution.

\subsection{State generation under the exact Hamiltonian}

To evaluate the performance of the approximate Hamiltonian $\hat{H}_{\text{app}}$ given in Eq.~(\ref{EffectiveHamiltonian}), in this section, we calculate the state generation under the exact Hamiltonian (\ref{ExactHamiltonian}) and the same initial state (\ref{InitialState}). In the next section, we will calculate the fidelity between the exact and approximate states. Fortunately, the exact states in this model can also been calculated analytically. Similar to the approximate solution case, we first diagonalize the exact Hamiltonian $\hat{H}_{\text{ext}}$. To this end, we introduce the transformation operator $\hat{T}(\hat{\lambda})=\exp[\hat{\lambda}(\hat{b}^{\dag}\hat{c}-\hat{c}^{\dag}\hat{b})]$ and displacement operators $\hat{D}_{b}(\hat{\zeta}_{b})=\exp[\hat{\zeta}_{b}(\hat{b}^{\dag}-\hat{b})]$ and $\hat{D}_{c}(\hat{\zeta}_{c})=\exp[\hat{\zeta}_{c}(\hat{c}^{\dag}-\hat{c})]$, where $\hat{\lambda}$, $\hat{\zeta}_{b}$, and $\hat{\zeta}_{c}$ are functions of $\hat{a}^{\dag }\hat{a}$. Then we can obtain the diagonalized exact Hamiltonian
\begin{eqnarray}
\hat{\tilde{H}}_{\text{ext}}&=&\hat{D}_{c}^{\dag}(\hat{\zeta}_{c})\hat{D}_{b}^{\dag}(\hat{\zeta}_{b})\hat{T}^{\dag}(\hat{\lambda})\hat{H}_{\text{ext}}\hat{T}(\hat{\lambda})\hat{D}_{b}(\hat{\zeta}_{b})\hat{D}_{c}(\hat{\zeta}_{c})\notag \\
&=&\hat{\tilde{H}}_{\text{ext}}^{a}+\hat{\tilde{H}}_{\text{ext}}^{b,c}+\hat{\tilde{H}}_{\text{ext}}^{a,b,c},
\end{eqnarray}
where the three parts $\hat{\tilde{H}}_{\text{ext}}^{a}$, $\hat{\tilde{H}}_{\text{ext}}^{b,c}$, and $\hat{\tilde{H}}_{\text{ext}}^{a,b,c}$ are defined by
\begin{subequations}
\label{diagonalizedexactHamiltonian3}
\begin{align}
\hat{\tilde{H}}_{\text{ext}}^{a}=&\tilde{\omega}_{a}\hat{a}^{\dag}\hat{a}+2g\hat{\zeta}_{b}\hat{a}^{\dag}\hat{a}(\xi_{c}\cos\hat{\lambda}-\xi_{b}\sin\hat{\lambda})\notag\\
&+2g\hat{\zeta}_{c}\hat{a}^{\dag }\hat{a}(\xi_{c}\sin\hat{\lambda}+\xi_{b}\cos\hat{\lambda})\notag\\
&+g\hat{a}^{\dag}\hat{a}(\hat{\zeta}_{c}^{2}-\hat{\zeta}_{b}^{2})\sin (2\hat{\lambda}),   \\
\hat{\tilde{H}}_{\text{ext}}^{b,c}=&(\hat{b}^{\dag}\hat{b}+\hat{\zeta}_{b}^{2})(\Delta_{b}\cos^{2}\hat{\lambda}+\Delta_{c}\sin^{2}\hat{\lambda})\notag\\
&+(\hat{c}^{\dag }\hat{c}+\hat{\zeta}_{c}^{2})(\Delta_{b}\sin ^{2}\hat{\lambda}+\Delta _{c}\cos ^{2}\hat{\lambda}),   \\
\hat{\tilde{H}}_{\text{ext}}^{a,b,c}=&g\hat{a}^{\dag }\hat{a}\hat{c}^{\dag}\hat{c}\sin (2\hat{\lambda})-g\hat{a}^{\dag }\hat{a}\hat{b}^{\dag}\hat{b}\sin (2\hat{\lambda}).
\end{align}
\end{subequations}
In Eq.~(\ref{diagonalizedexactHamiltonian3}) we introduced the photon-number-dependent rotation angle
\begin{eqnarray}
\hat{\lambda}(\hat{a}^{\dag }\hat{a})=\frac{1}{2}\arctan\left(\frac{2g\hat{a}^{\dag}\hat{a}}{\Delta_{c}-\Delta_{b}}\right),
\end{eqnarray}
and the two photon-number-dependent displacement amplitudes are
\begin{subequations}
\begin{align}
\hat{\zeta}_{b}(\hat{a}^{\dag }\hat{a})=\frac{g\hat{a}^{\dag}\hat{a}(\xi_{c}\cos\hat{\lambda}-\xi_{b} \sin \hat{\lambda})}{g\hat{a}^{\dag }\hat{a}\sin (2\hat{\lambda})-(\Delta_{b}\cos ^{2}\hat{\lambda}+\Delta_{c}\sin^{2}\hat{\lambda})},\\
\hat{\zeta}_{c}(\hat{a}^{\dag }\hat{a})=-\frac{g\hat{a}^{\dag}\hat{a}(\xi_{c}\sin\hat{\lambda}+\xi_{b}\cos\hat{\lambda})}{g\hat{a}^{\dag }\hat{a}\sin (2\hat{\lambda})+(\Delta _{b}\sin^{2}\hat{\lambda}+\Delta_{c}\cos ^{2}\hat{\lambda})}.
\end{align}
\end{subequations}

Based on the diagonalized exact Hamiltonian $\hat{\tilde{H}}_{\text{ext}}$, we can obtain the exact evolution of the system starting from the initial state ~(\ref{InitialState}). The exact evolution state can be obtained as
\begin{eqnarray}
\left\vert\Psi(t)\right\rangle&=&\text{e}^{-i\hat{H}_{\text{ext}}t}\left\vert\psi(0)\right\rangle \notag \\
&=& \frac{1}{\sqrt{2}}[\left\vert 0\right\rangle _{a}\left\vert 0 \right\rangle_{b}\left\vert 0 \right\rangle_{c}+e^{-i\theta_{2}(t)}\left\vert 1\right\rangle _{a}\left\vert \alpha_{2}(t)\right\rangle _{b}\left\vert \beta_{2}(t)\right\rangle _{c}],\nonumber \\ \label{Psi(t)}
\end{eqnarray}
where we introduce the phase
\begin{equation}
\theta_{2}(t)=\tilde{\omega}_{a}t+\Lambda_{2}t+\phi_{2}(t),
\end{equation}
with
\begin{subequations}
\begin{align}
\Lambda_{2}=&\zeta_{b}^{2}(1)[\Delta_{b}\cos^{2}\lambda(1)+\Delta_{c}\sin^{2}\lambda(1)]\notag \\
&+\zeta_{c}^{2}(1)[\Delta_{b}\sin^{2}\lambda(1)+\Delta_{c}\cos^{2}\lambda(1)] \notag \\
&+2g\zeta_{b}(1)[\xi_{c}\cos\lambda(1)-\xi_{b}\sin\lambda(1)]\notag \\
&+2g\zeta_{c}(1)[\xi_{c}\sin\lambda(1)-\xi_{b}\cos\lambda(1)]\notag \\
&+g[\zeta_{c}^{2}(1)-\zeta_{b}^{2}(1)]\sin2\lambda(1),\\
\phi_{2}(t)=&\zeta_{c}^{2}(1)\text{sin}(\Lambda_{c}t)-\zeta_{b}^{2}(1)\text{sin}(\Lambda_{b}t).
\end{align}
\end{subequations}
The coherent-state amplitudes in Eq.~(\ref{Psi(t)}) are given by
\begin{subequations}
\begin{align}
\alpha_{2}(t)=&\zeta_{b}(1)(1-e^{i\Lambda_{b}t})\text{cos}\lambda(1)+\zeta_{c}(1)(1-e^{-i\Lambda_{b}t})\text{sin}\lambda(1),\\
\beta_{2}(t)=&\zeta_{c}(1)(1-e^{-i\Lambda_{c}t})\text{cos}\lambda(1)-\zeta_{b}(1)(1-e^{i\Lambda_{b}t})\text{sin}\lambda(1),
\end{align}
\end{subequations}
where
\begin{subequations}
\begin{align}
\Lambda_{b}=g\sin2\lambda(1)-\Delta_{b}\cos^{2}\lambda(1)-\Delta_{c}\sin^{2}\lambda(1),\\
\Lambda_{c}=g\sin2\lambda(1)+\Delta_{b}\sin^{2}\lambda(1)+\Delta_{c}\cos^{2}\lambda(1).
\end{align}
\end{subequations}

We also express the states of mode $a$ with the bases $|\pm\rangle=(\left\vert0\right\rangle_{a}\pm\left\vert1\right\rangle_{a}) /\sqrt{2}$. Then the generated state $\left\vert\Psi(t)\right\rangle$ can be expressed as
\begin{equation}
|\Psi(t)\rangle=\frac{1}{2\mathcal{K}_{+}(t)}|+\rangle\left|\Psi_{+}(t)\right\rangle+\frac{1}{2\mathcal{K}_{-}(t)}|-\rangle\left|\Psi_{-}(t)\right\rangle,
\end{equation}
where we introduce the exact entangled cat states for the two modes as
\begin{equation}
\left\vert\Psi_{\pm}(t)\right\rangle=\mathcal{K}_{\pm}(t)\left( \left\vert 0 \right\rangle_{b}\left\vert 0 \right\rangle_{c}\pm e^{-i\theta_{2}(t)}\left\vert \alpha_{2}(t)\right\rangle_{b}\left\vert \beta_{2}(t)\right\rangle_{c}\right),\label{ExactEntangledCoherentStates}
\end{equation}
with the normalization constants
\begin{equation}
\mathcal{K}_{\pm}(t)=\left(2\{1\pm e^{-\frac{1}{2}(|\alpha_{2}(t)|^{2}+|\beta_{2}(t)|^{2})}\text{cos}[\theta_{2}(t)] \}\right)^{-1/2}.\label{NormalizationConstantsK}
\end{equation}
By performing a measurement of mode $a$ in the basis states $\left\vert\pm \right\rangle$, the two bosonic modes $b$ and $c$ will collapse into the entangled cat states $\left\vert\Psi_{\pm }(t)\right\rangle.$
The corresponding detection probabilities are
\begin{equation}
P_{\pm}(t)=\frac{1}{4\left\vert \mathcal{K}_{\pm}(t)\right\vert ^{2}},\label{ExactProbabilities}
\end{equation}
which represent the success probabilities for generation of entangled cat states.

\begin{figure}[tbp]
\centering
\includegraphics[scale=0.69]{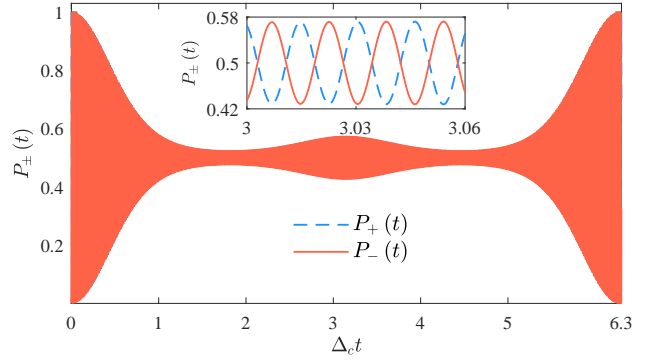}
\caption{Plot of the exact detection probabilities $ P_{\pm}(t)$ given in Eq.~(\ref{ExactProbabilities}) as functions of the evolution time $\Delta_{c}t$. The inset shows a close-up plot of the probabilities in the middle duration of one period. The other parameters are $\omega_{a}/\Delta_{c}=1.1$, $g/\Delta_{c}=0.01$, $\Delta_{b}/\Delta_{c}=2$, and $\Omega_{b}/\Delta_{c}=\Omega_{c}/\Delta_{c}=200$.}
\label{Detprob(closed)}
\end{figure}

In Fig.~\ref{Detprob(closed)} we show the time evolution of the measurement probabilities $P_{\pm}(t)$ defined in Eq.~(\ref{ExactProbabilities}). Figure~\ref{Detprob(closed)} shows that the detection probabilities exhibit large magnitude oscillations around the two ends in one period, while the magnitude of oscillations decreases to about $\frac{1}{2}$ in the middle duration of one period. It can also be seen from Eq.~(\ref{NormalizationConstantsK}) that, at the detection time $t_{s}$, we have $\left\vert \mathcal{K}_{\pm}(t_{s})\right\vert ^{2}\approx \frac{1}{2}$ and then the probabilities $P_{\pm}(t_{s})\approx \frac{1}{2}$. In addition, we analyze the time evolution of the approximate measurement probabilities $\mathcal{P}_{\pm}(t)$ given by Eq.~(\ref{EffectiveProbabilities}) and find that the approximate results (both the whole envelope and the details of the oscillation) are in good agreement with the exact results.

\subsection{Fidelities between the approximate and exact states in the closed-system case\label{idilitysec}}

The validity of the approximate Hamiltonian $\hat{H}_{\text{app}}$ can be evaluated by checking the fidelity $F(t)=|\langle\Psi(t)|\psi(t)\rangle|^{2}$ between the approximate state $|\psi(t)\rangle$ and the exact state $|\Psi(t)\rangle$, which are defined in Eqs.~(\ref{psi(t)}) and ~(\ref{Psi(t)}), respectively. Then the expression of the fidelity can be obtained as
\begin{equation}
F(t)=\frac{1}{4} \left|1+e^{-i[\theta_{1}(t)+\theta_{2}(t)]-(|\alpha_{1}|^{2}+|\beta_{1}|^{2}+|\alpha_{2}|^{2}+|\beta_{2}|^{2})/2+\alpha_{1}\alpha_{2}^{\ast}+\beta_{1}\beta_{2}^{\ast}}\right|^{2}.\label{Ft}
\end{equation}

Similarly, we also evaluate the performance of the entangled cat state by calculating the fidelities between the generated states $|\Psi_{\pm}(t)\rangle$ (after the measurement) in Eq.~(\ref{ExactEntangledCoherentStates}) and the target states $|\psi_{\pm}(t)\rangle$ (the entangled cat states) in Eq.~(\ref{EffectiveEntangledCoherentStates}). Using Eqs.~(\ref{EffectiveEntangledCoherentStates}) and (\ref{ExactEntangledCoherentStates}), the fidelities $F_{\pm}(t)=|\langle\Psi_{\pm}(t)\mid\psi_{\pm}(t)\rangle|^{2}$ can be obtained as
\begin{eqnarray}
F_{\pm}(t)&=&|\mathcal{M}_{\pm}|^{2}|\mathcal{K}_{\pm}|^{2} \left|1 \pm e^{i\theta_{1}(t)-(|\alpha_{1}|^{2}+|\beta_{1}|^{2})/2}\pm e^{-i\theta_{2}(t)-(|\alpha_{2}|^{2}+|\beta_{2}|^{2})/2}  \right.\notag \\
&&\left.+ e^{i[\theta_{1}(t)-\theta_{2}(t)]-(|\alpha_{1}|^{2}+|\beta_{1}|^{2}+|\alpha_{2}|^{2}+|\beta_{2}|^{2}+\alpha_{1}^{\ast}\alpha_{2}^{\ast}+\beta_{1}^{\ast}\beta_{2}^{\ast})/2} \right|^{2}.
\end{eqnarray}

\begin{figure}[tbp]
\centering
\includegraphics[scale=0.75]{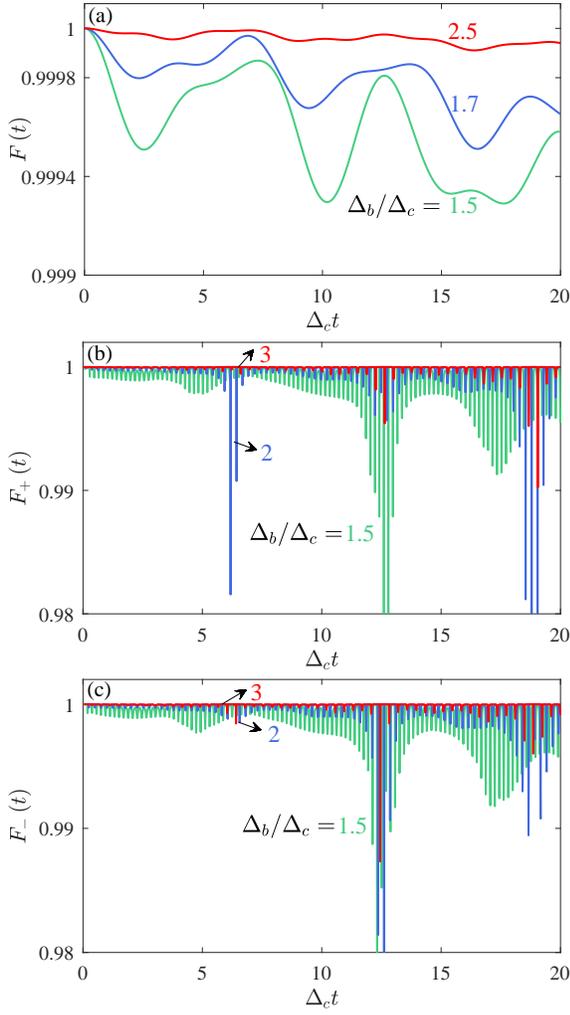}
\caption{Fidelities $F(t)$ and $F_{\pm}(t)$ versus the evolution time $\Delta_{c}t$. The curves correspond to the values (a) $\Delta_{b}/\Delta_{c}=1.5$ (green), $\Delta_{b}/\Delta_{c}=1.7$ (blue), and $\Delta_{b}/\Delta_{c}=2.5$ (red) and (b) and (c) $\Delta_{b}/\Delta_{c}=1.5$ (green), $\Delta_{b}/\Delta_{c}=2.0$ (blue), and $\Delta_{b}/\Delta_{c}=3.0$ (red). The other parameters used in panels(a-c) are $g/\Delta_{c}=0.01$, $\Omega_{b}/\Delta_{c}=\Omega_{c}/\Delta_{c} =50$, and $\omega_{a}/\Delta_{c}=0.1$.}
\label{Fidelity(closed)}
\end{figure}

In Fig.~\ref{Fidelity(closed)} we plot the fidelities $F(t)$ and $F_{\pm}(t)$ as functions of time $t$ when the ratio $\Delta_{b}/\Delta_{c}$ takes different values. Here we can see that the fidelities exhibit fast oscillation because of the high-frequency oscillation terms $\text{exp}[i\theta_{1}(t)]$ and $\text{exp}[i\theta_{2}(t)]$. Note that in our simulations, we choose $\Delta_{b}/\Delta_{c}$ as an adjustable parameter, because when we consider the few-excitation conditions, the approximate conditions also can be written as $|\xi_{c}|\gg 1/\left\vert\ 1-\Delta_{c}/\Delta_{b}\right\vert$ and $|\xi_{b}|\gg 1 /\left\vert\Delta_{b}/\Delta_{c}-1\right\vert$, so we use $\Delta_{b}/\Delta_{c}$ as a critical parameter for investigating fidelities. In addition, the envelope of the fidelity is larger for a larger value of $\Delta_{b}/\Delta_{c}$, which means the parameter conditions of our approximation are satisfied when $\Delta_{b}/\Delta_{c}\geq2$.

\section{The open-system case \label{opensystem}}

In this section we study the generation of entangled cat states in the open-system case. In particular, we analyze the influence of the dissipations on the fidelity, the success probability, and the degree of entanglement of the generated states.

\subsection{Quantum master equation}

To include the damping and noise effects in this system, we assume that the three bosonic modes are coupled to three individual Markovian environments. The evolution of the system can be described by the quantum master equation
\begin{eqnarray}
\dot{\hat{\rho}}&=&i[\hat{\rho}, \hat{H}_{I}(t)]+\kappa _{a}(\bar{n}_{a}+1)\mathcal{D}[\hat{a}]\hat{\rho}+\kappa _{a}\bar{n}_{a}\mathcal{D}[\hat{a}^{\dagger }]\hat{\rho} \notag \\
&&+\kappa_{b}( \bar{n}_{b}+1) \mathcal{D}[\hat{b}]\hat{\rho} +\kappa _{b}\bar{n}_{b}\mathcal{D}[\hat{b}^{\dagger }]\hat{\rho} \notag \\
&&+\kappa_{c}( \bar{n}_{c}+1) \mathcal{D}[\hat{c}]\hat{\rho} +\kappa _{c}\bar{n}_{c}\mathcal{D}[\hat{c}^{\dagger }]\hat{\rho},
\label{mastereqschopic}
\end{eqnarray}
where the Hamiltonian $\hat{H}_{I}(t)$ is given by Eq.~(\ref{rotatedHamiltonian}), $\mathcal{D}[\hat{o}=\hat{a}, \hat{a}^{\dagger}, \hat{b}, \hat{b}^{\dagger}, \hat{c}, \hat{c}^{\dagger}]\hat{\rho} =\hat{o}\hat{\rho} \hat{o}^{\dag }-(\hat{o}^{\dag }\hat{o}\hat{\rho} +\hat{\rho} \hat{o}^{\dag }\hat{o})/2$ is the
standard Lindblad superoperator that describes the damping of these bosonic modes, the parameters $\kappa_{a}$ ($\kappa _{b}$ and $\kappa _{c}$) and $\bar{n}_{a}$ ( $\bar{n}_{b}$ and $\bar{n}_{c}$) are, respectively, the damping rates and the environment thermal excitation occupations of the optical mode $a$ ($b$ and $c$).

Similar to the closed-system case, we perform the displacement transformations to the quantum master equation by
\begin{equation}
\hat{\rho} ^{(1)}=\hat{D}_{c}^{\dag }[\chi_{c}(t)] \hat{D}_{b}^{\dag }[\chi_{b}(t)] \hat{\rho} \hat{D}_{b}[\chi_{b}(t)] \hat{D}_{c}[\chi_{c}(t)],
\end{equation}
where $\hat{\rho}^{(1)}$ is the density matrix of the three-mode system in the displacement representation, $\hat{D}_{b}[\chi_{b}(t)]$ and $\hat{D}_{c}[\chi_{c}(t)]$ are the displacement operators, and $\chi_{b}(t)$ and $\chi_{c}(t) $ are the coherent displacement amplitudes, which need to be determined in the transformed master equation. By substituting the relation $\hat{\rho}=\hat{D}_{b}[\chi_{b}(t)] \hat{D}_{c}[\chi_{c}(t)]\hat{\rho} ^{(1)}\hat{D}_{c}^{\dag }[\chi_{c}(t)] \hat{D}_{b}^{\dag }[\chi_{b}(t)]$ into Eq.~(\ref{mastereqschopic}) and using the differential of these displacement operators with respect to time $t$, we can obtain the quantum master equation in the displacement representation as
\begin{eqnarray}
\frac{d}{dt}\hat{\rho} ^{(1)}&=&i[\hat{\rho}^{(1)},\hat{H}^{(1)}(t)]+\kappa _{a}\left( \bar{n}_{a}+1\right) \mathcal{D}[\hat{a}]\hat{\rho}^{(1)}+\kappa _{a}\bar{n}_{a}\mathcal{D}[\hat{a}^{\dagger }]\hat{\rho}^{(1)}  \notag \\
&&+\kappa _{b}( \bar{n}_{b}+1) \mathcal{D}[\hat{b}]\hat{\rho}^{(1)}+\kappa _{b}\bar{n}_{b}\mathcal{D}[\hat{b}^{\dagger }]\hat{\rho}^{(1)}  \notag \\
&&+\kappa _{c}( \bar{n}_{c}+1) \mathcal{D}[\hat{c}]\hat{\rho}^{(1)}+\kappa _{c}\bar{n}_{c}\mathcal{D}[\hat{c}^{\dagger }]\hat{\rho}^{(1)},\label{MasterEquation}
\end{eqnarray}
where the transformed Hamiltonian takes the form
\begin{eqnarray}
\hat{H}^{(1)}&=&\{\omega _{c}+g[ \chi_{b}^{\ast}(t)\chi_{c}(t)+\chi_{c}^{\ast }(t)\chi_{b}(t)]\} \hat{a}^{\dag }\hat{a}  \notag \\
&&+g\hat{a}^{\dag }\hat{a}(\hat{b}^{\dag }\hat{c}+\hat{c}^{\dag }\hat{b})+\Delta _{b}\hat{b}^{\dag }\hat{b}+\Delta _{c}\hat{c}^{\dag }\hat{c}  \notag \\
&&+g\hat{a}^{\dag }\hat{a}[\chi_{c}(t)\hat{b}^{\dag} +\chi_{b}^{\ast }(t)\hat{c}+\text{H.c.}].
\end{eqnarray}
The coherent displacement amplitudes $\chi_{b}(t) $ and $\chi_{c}(t)$ are governed by
\begin{subequations}
\begin{align}
\dot{\chi}_{b}(t)=-i\Omega_{b}-i\Delta_{b}\chi_{b}(t)-\kappa_{b}\chi_{b}(t)/2,\\
\dot{\chi}_{c}(t)=-i\Omega_{c}-i\Delta_{c}\chi_{c}(t)-\kappa_{c}\chi_{c}(t)/2,
\end{align}
\end{subequations}
which are determined by eliminating the displacement terms in the master equation.

We consider the steady-state displacement case, in which the time scale of the system approaching its steady state is much shorter than other evolution time scales. In this case, we have $\chi_{b,\text{ss}}=\Omega _{b}/( i\kappa_{b}/2-\Delta _{b})$ and $\chi_{c,\text{ss}}=\Omega_{c}/(i\kappa_{c}/2-\Delta_{c})$, which are controllable complex numbers by selecting $\Omega_{b}$ and $\Omega _{c}$. The Hamiltonian $\hat{H}^{(1)}$ becomes
\begin{eqnarray}
\hat{H}_{\text{ext}}^{(1)}&=&\omega _{a}^{(1)}\hat{a}^{\dag }\hat{a}+\Delta _{b}\hat{b}^{\dag }\hat{b}+\Delta _{c}\hat{c}^{\dag }\hat{c}+g\hat{a}^{\dag }\hat{a}(\hat{b}^{\dag }\hat{c}+\hat{c}^{\dag }\hat{b})\notag \\
&&+g\hat{a}^{\dag }\hat{a}(\chi_{c,\text{ss}}\hat{b}^{\dag }+\chi_{c,\text{ss}}^{\ast}\hat{b})+g\hat{a}^{\dag }\hat{a}(\chi_{b,\text{ss}}\hat{c}^{\dag}+\chi_{b,\text{ss}}^{\ast }\hat{c}), \nonumber \\ \label{OpenExactHamiltonian}
\end{eqnarray}
where $\omega_{a}^{(1)}=\omega_{c}+g(\chi_{b,\text{ss}}^{\ast}\chi_{c,\text{ss}}+\chi_{c,\text{ss}}^{\ast}\chi_{b,\text{ss}})$. Under the parameter conditions $|\chi_{c,\text{ss}}|\gg |\Delta _{b}| \sqrt{n_{c}}/| \Delta
_{b}-\Delta _{c}|$ and $|\chi_{b,\text{ss}}|\gg |\Delta_{c}| \sqrt{n_{b}}/| \Delta _{b}-\Delta _{c}|$, the term $g\hat{a}^{\dag}\hat{a}(\hat{b}^{\dag }\hat{c}+\hat{c}^{\dag }\hat{b})$ in Eq.~(\ref{OpenExactHamiltonian}) is safely ignored, and we obtain
\begin{eqnarray}
\hat{H}_{\text{app}}^{(1)}&=&\omega_{a}^{(1)}\hat{a}^{\dag }\hat{a}+\Delta _{b}\hat{b}^{\dag }\hat{b}+\Delta _{c}\hat{c}^{\dag }\hat{c}+g\hat{a}^{\dag }\hat{a}(\chi_{c,\text{ss}}\hat{b}^{\dag}+\chi_{c,\text{ss}}^{\ast}\hat{b})\notag \\
&&+g\hat{a}^{\dag }\hat{a}(\chi_{b,\text{ss}}\hat{c}^{\dag}+\chi_{b,\text{ss}}^{\ast }\hat{c}).\label{OpenEffectiveHamiltonian}
\end{eqnarray}

For our state generation motivation, we consider the initial state $|\pm\rangle\left\vert 0\right\rangle _{b}\left\vert 0\right\rangle _{c}$. The evolution of the density matrix of the system can be obtained by numerically solving the quantum master equation~(\ref{MasterEquation}), and then the properties of the generated states can be calculated.

\begin{figure}[tbp]
\centering
\includegraphics[scale=0.44]{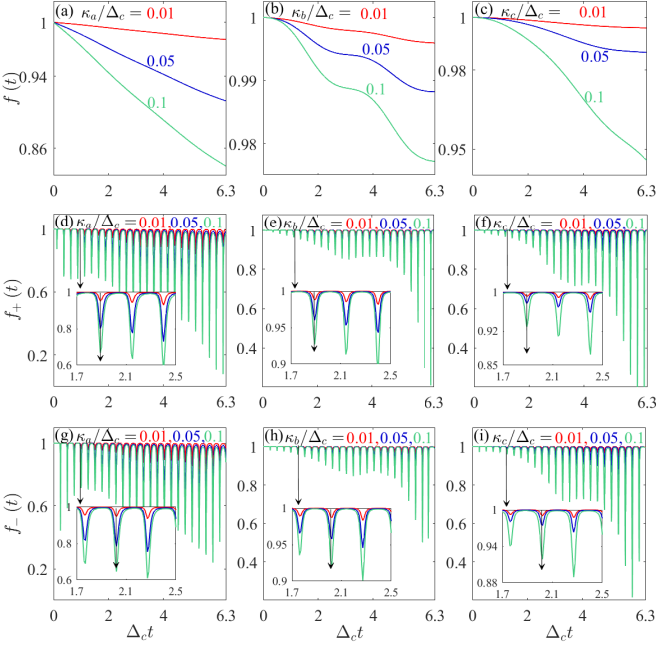}
\caption{Fidelities $f(t)$ and $f_{\pm}(t)$ as functions of the evolution time $\Delta_{c}t$ in various cases: (a), (d), and (g) $\kappa_{b}/\Delta _{c}=\kappa _{c}/\Delta _{c}=0.001$ and $\kappa_{a}/\Delta_{c}=0.01, 0.05,$ and $0.1$; (b), (e), and (h) $\kappa_{a}/\Delta_{c}=\kappa_{c}/\Delta_{c}=0.001$ and $\kappa_{b}/\Delta_{c}=0.01,0.05,$ and $0.1$; and (c), (f), and (i) $\kappa_{a}/\Delta_{c}=\kappa_{b}/\Delta_{c}=0.001$ and $\kappa_{c}/\Delta_{c}=0.01,0.05,$ and $0.1$. The insets show zoomed-in plots of the details of fidelities over the duration $\Delta_{c}t=1.7$-$2.5$. The other parameters are $\omega_{a}/\Delta_{c}=0.1$, $g/\Delta_{c}=0.01$, $\Delta_{b}/\Delta_{c}=2$, and $\bar{n}_{a}=\bar{n}_{b}=\bar{n}_{c}=0.$}
\label{Fidelity(open)}
\end{figure}

\subsection{Fidelities between the approximate and the exact states in the open-system case}

To study the influence of dissipation of the system on the state generation, we calculate the fidelity between the exact density matrix $\hat{\rho}^{\text{ext}}(t)$ and the approximate state $|\psi(t)\rangle$ in Eq.~(\ref{psi(t)}) as
\begin{eqnarray}
f(t)&=&\langle\psi(t)|\hat{\rho}^{\text{ext}}(t)|\psi(t)\rangle.
\end{eqnarray}
For generation of two-mode entangled cat states, we perform a measurement of mode $a$ in the bases $|\pm\rangle$, and then the corresponding density matrices become
\begin{eqnarray}
\hat{\rho}_{\text{ext}}^{(\pm)}(t)=\frac{1}{P_{\pm}^{\text{ext}}(t)}\left\langle\pm\right\vert\hat{\rho}^{\text{ext}}(t)\left\vert \pm \right\rangle,
\end{eqnarray}
where $P_{\pm}^{\text{ext}}(t)=\text{Tr}[\left\langle\pm\right\vert\hat{\rho}^{\text{ext}}(t)\left\vert\pm\right\rangle]$ are the probabilities for detecting the states $|\pm\rangle$ of mode $a$. Accordingly, the fidelities between the generated states $\hat{\rho}_{\text{ext}}^{(\pm)}(t)$ and the target states $|\psi_{\pm}(t)\rangle$ can be calculated as
\begin{eqnarray}
f_{\pm}(t)&=&\langle\psi_{\pm}(t)|\hat{\rho}_{\text{ext}}^{(\pm)}(t)|\psi_{\pm}(t)\rangle .
\end{eqnarray}
\begin{figure}[tbp]
\centering
\includegraphics[scale=0.55]{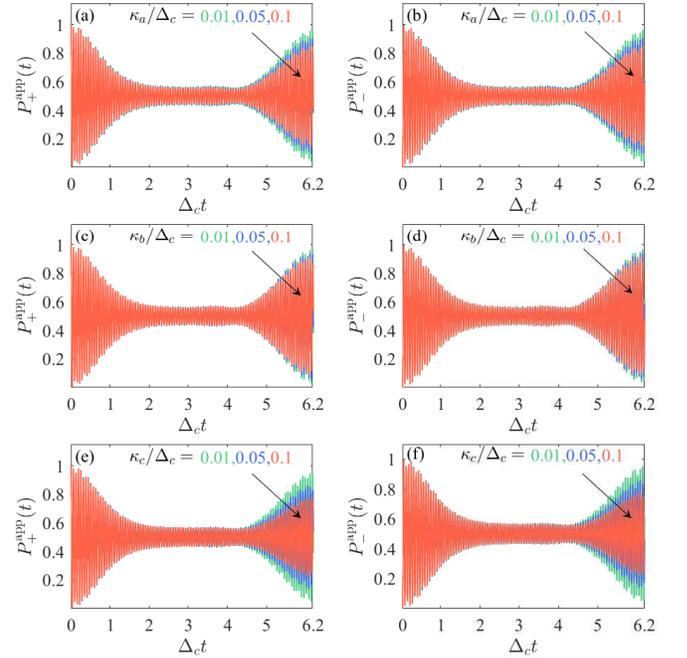}
\caption{Probabilities $P_{\pm}^{\text{app}}(t)$ as functions of the evolution time $\Delta_{c}t$ in various cases: (a) and (b) $\kappa_{b}/\Delta _{c}=\kappa _{c}/\Delta _{c}=0.001$ and $\kappa_{a}/\Delta_{c}=0.01$, $0.05$, and $0.1$; (c) and (d) $\kappa_{a}/\Delta_{c}=\kappa_{c}/\Delta_{c}=0.001$ and $\kappa_{b}/\Delta_{c}=0.01$, $0.05$, and $0.1$; and (e) and (f) $\kappa_{a}/\Delta_{c}=\kappa_{b}/\Delta_{c}=0.001$ and $\kappa_{c}/\Delta_{c}=0.01$, $0.05$, and $0.1$. The other parameters are $\omega_{a}/\Delta_{c}=0.1$, $g/\Delta_{c}=0.02$, $\Delta_{b}/\Delta_{c}=2.0$, and $\bar{n}_{a}$=$\bar{n}_{b}$=$\bar{n}_{b}=0$.}
\label{DetProb(open)}
\end{figure}

In Fig.~\ref{Fidelity(open)} we plot the fidelities $f(t)$ and $f_{\pm}(t)$ versus the evolution time $\Delta_{c}t$ at different values of the scaled decay rate $\kappa_{a}/\Delta_{c}$, $\kappa_{b}/\Delta_{c}$, and $\kappa_{c}/\Delta_{c}$. In Figs.~\ref{Fidelity(open)}(a)-(c) we can see that as the system evolves, the fidelity undergoes a decay dynamics. The curve of the fidelity $f(t)$ decays with the evolution of the system gently. In Figs.~\ref{Fidelity(open)}(d)-(i) we show how $f_{\pm}(t)$ evolve at various values of the decay rates $\kappa_{a}/\Delta_{c}$, $\kappa_{b}/\Delta_{c}$, and $\kappa_{c}/\Delta_{c}$. We can see that fidelities experience fast oscillation. In addition, the dissipation rates $\kappa_{a}/\Delta_{c}$, $\kappa_{b}/\Delta_{c}$, and $\kappa_{c}/\Delta_{c}$ have a greater impact on the curve of $f_{\pm}(t)$, which is similar to the behavior of the fidelity $f(t)$. The greater the decay rates, the lower the envelope of these fidelities, as shown in the insets. Note that the scaled decay rates used are within the reach of current experimental conditions (see Sec.~\ref{Discussion} for a detailed analysis).

\subsection{Detection probabilities}

Similar to the closed-system case, for generation of entangled cat states, we perform the projective measurement of mode $a$ on the states $\left\vert \pm \right\rangle=\left( \left\vert 0\right\rangle _{a}\pm \left\vert 1\right\rangle_{a}\right) /\sqrt{2}$, and the reduced density matrices of modes $b$ and $c$ become
\begin{eqnarray}
\hat{\rho}_{\text{app}}^{(\pm )}(t)=\frac{1}{P_{\pm}^{\text{app}}(t)}\left\langle\pm\right\vert\hat{\rho}^{\text{app}}(t)\left\vert \pm \right\rangle,
\end{eqnarray}
where $P_{\pm}^{\text{app}}(t)=\text{Tr}[\left\langle\pm\right\vert\hat{\rho}^{\text{app}}(t)\left\vert \pm \right\rangle]$ are the corresponding probabilities for detecting the states $\left\vert \pm\right\rangle _{a}$.

To analyze the state generation probabilities in the open-system case, we plot the time evolution of the probabilities $P_{\pm}^{\text{app}}(t)$ at selected values of the scaled decay rates $\kappa_{a}/\Delta_{c}$, $\kappa_{b}/\Delta_{c}$, and $\kappa_{c}/\Delta_{c}$. Figures ~\ref{DetProb(open)}(a)-\ref{DetProb(open)}(d) show that the probabilities $P_{\pm}^{\text{app}}(t)$ oscillate rapidly at the beginning, and then the amplitude of the oscillation envelope decreases gradually with the evolution of the system. In the middle duration of $\Delta_{c}t\approx2.0-4.5$, the value of the probabilities $P_{+}^{\text{app}}(t)\approx P_{-}^{\text{app}}(t)\rightarrow 1/2$. However, in the latter part of a period, the oscillation amplitude envelope will revive, and the amplitude of the revival envelope is smaller for a larger value of the dissipation rates. We also see that the greater the dissipation rate, the smaller the amplitude of the oscillation.

\begin{figure}[tbp]
\centering
\includegraphics[scale=0.55]{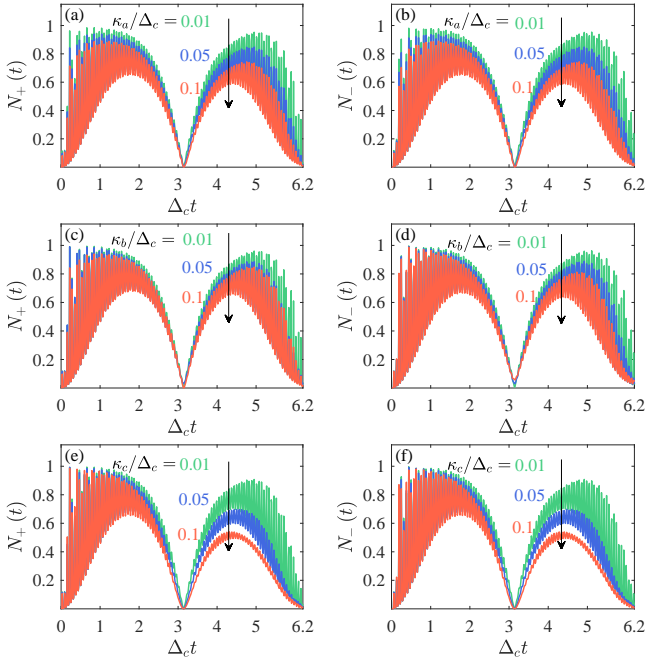}
\caption{Logarithmic negativities $N_{\pm }(t) $ as functions of the scaled evolution time $\Delta _{c}t$ in various cases: (a) and (b) $\kappa _{b}/\Delta _{c}=\kappa _{c}/\Delta _{c}=0.001$ and $\kappa _{a}/\Delta_{c}=0.01$, $0.05$, and $0.1$; (c) and (d) $\kappa _{a}/\Delta _{c}=\kappa_{c}/\Delta _{c}=0.001$ and $\kappa _{b}/\Delta _{c}=0.01$, $0.05$, and $0.1$; and (e) and (f) $\kappa_{a}/\Delta_{c}=\kappa_{b}/\Delta_{c}=0.001$ and $\kappa_{c}/\Delta_{c}=0.01$, $0.05$, and $0.1$. The other parameters are $\omega_{a}/\Delta_{c}=0.1$, $g/\Delta_{c}=0.02$, $\Delta_{b}/\Delta_{c}=2.0$, and $\bar{n}_{a}$=$\bar{n}_{b}$=$\bar{n}_{c}=0$.}
\label{Entanglement(open)}
\end{figure}

\subsection{Entanglement dynamics of the generated cat states}

In the open-system case, we use the logarithmic negativity to describe the quantum entanglement between the two mechanical-like modes ($b$ and $c$) for the generated density matrices $\hat{\rho}_{\text{app}}^{(\pm )}(t)$. In terms of Eqs.~(\ref{LogarithmicNegativity}), ~(\ref{MasterEquation}), and~(\ref{OpenEffectiveHamiltonian}), we can numerically solve the logarithmic negativity of the states $\hat{\rho}_{\text{app}}^{(\pm )}(t)$. In Fig.~\ref{Entanglement(open)}, we plot the time evolution of the logarithmic negativity when these dissipation rates of the system take different values. Here we can see, with the evolution of the system, that the logarithmic negativity oscillates very fast. In our simulations, we find that at $\Delta_{c}t=n\pi$, for natural numbers $n$, the cavity mode and the mechanical-like modes are decoupled from each other and the logarithmic negativity $N_{\pm}(t)$ is close to zero. At each cycle, with the increase of time $\Delta_{c}t$, the logarithmic negativity first increases rapidly. In the middle duration of $\Delta_{c}t\approx1-2$, the logarithmic negativity reaches the maximum. At the end of one duration, the logarithmic negativity decreases gradually. In addition, we show the influence of the decay rates on the logarithmic negativity. As the decay rate increases, the maximum value of the logarithmic negativity decreases, which means that the logarithmic negativity decays faster for larger decay rates.

\begin{table*}[t]
\centering
\caption{Parameters used in our numerical simulations: the driving detunings (the effective resonance frequencies of the mechanical-like modes in the displacement representation) $\Delta_{b}=\omega_{b}-\omega_{L}$ and $\Delta_{c}=\omega_{c}-\omega_{L}$ of modes $b$ and $c$, the Fredkin-type interaction strength $g$, the displacement amplitudes $|\xi_{b}|=\Omega_{b}/|\Delta_{b}|$ and $|\xi_{c}|=\Omega_{c}/|\Delta_{c}|$, the enhanced single-photon optomechanical-coupling strengths $g_{b}=g|\xi_{b}|$ and $g_{c}=g|\xi_{c}|$, and the decay rates $\kappa_{a}$, $\kappa_{b}$, and $\kappa_{c}$ of modes $a$, $b$, and $c$.}
\label{table1}
\begin{tabular*}{1\textwidth}{@{\extracolsep{\fill}}c c c c}
\toprule
Symbol & Remark & Scaled parameter  &Parameter  \\

\hline
$\Delta_{c}$ &  chosen as the frequency scale  & $1$ & $2\pi\times 10$ MHz \\
\hline
$\Delta_{b}$ & effective frequency of mode $b$  & $\Delta_{b}/\Delta_{c}=1$ - $10$ & $2\pi\times$ ($10$ - $100$) MHz \\
\hline
$g$   & Fredkin-type interaction strength &  $g/\Delta_{c}=0.01$ & $2\pi\times$ $100$ kHz \\
\hline
$|\xi_{b}|$ & $|\xi_{b}|\gg1$  &  & $50$ - $200$  \\
\hline
$|\xi_{c}|$ & $|\xi_{c}|\gg1$  &  & $50$ - $200$  \\
\hline
$g_{b}=g|\xi_{b}|$ & enhanced OM-coupling strength  & $g_{b}/\Delta_{c}=0.5$ - $2$   & $2\pi\times$ ($5$ - $20$) MHz  \\
\hline
$g_{c}=g|\xi_{c}|$ & enhanced OM-coupling strength  &  $g_{c}/\Delta_{c}=0.5$ - $2$ & $2\pi\times$ ($5$ - $20$) MHz  \\
\hline
$\kappa_{a}$  & decay rate of mode $a$  & $\kappa_{a}/\Delta_{c}=0.001$ - $0.1$   &  $2\pi\times$ ($10$ - $1000$) kHz \\
\hline
$\kappa_{b}$  & decay rate of mode $b$  & $\kappa_{b}/\Delta_{c}=0.001$ - $0.1$   &  $2\pi\times$ ($10$ - $1000$) kHz \\
\hline
$\kappa_{c}$  & decay rate of mode $c$  & $\kappa_{c}/\Delta_{c}=0.001$ - $0.1$   &  $2\pi\times$ ($10$ - $1000$) kHz \\
\hline

\botrule
\end{tabular*}
\end{table*}

\section{Discussions on the experimental implementation} \label{Discussion}

In this section we present a discussion of the experimental implementation of the present scheme. The physical model considered in this work consists of a Fredkin-type interaction involving three bosonic modes $a$, $b$, and $c$. Here modes $b$ and $c$ are strongly driven by two laser fields with the same driving frequency $\omega_{L}$ and individual driving amplitudes $\Omega_{b}$ and $\Omega_{c}$. It should be pointed out that the present physical model is general and hence it can be implemented with various physical platforms in which the Fredkin-type interaction and the bosonic drivings can be realized.
Usually, the strong drivings could be reliably realized in many systems, and both the driving amplitudes and frequency are adjustable parameters. Then the key point for the experimental implementation of this scheme is the realization of the Fredkin-type interaction involving three bosonic modes. Fortunately, the Fredkin-type interaction was experimentally realized in a circuit-QED system recently~\cite{GAO2019NATURE}. Hence, the parameters given by these experiments provide us some references. It should be pointed out that the parameters we used in our simulations are of the same order as the reported parameters; these choices will ensure that our parameters are experimentally accessible.

In the circuit-QED system, the Fredkin-type interaction involving three microwave fields was realized by using two beam-splitter transformations and a cross-Kerr interaction~\cite{GAO2019NATURE}. In particular, the interaction strength of the obtained Fredkin interaction is half of the cross-Kerr interaction strength. In this system, the real experimental parameters are
$\omega_{a}=2\pi\times8.493$ GHz, $\omega_{b}=2\pi\times9.32$ GHz, $\omega_{c}=2\pi\times7.249$ GHz, $\chi=2\pi\times2.59\times10^{3}$ kHz, $\kappa_{a}=2\pi\times1.25$ kHz, $\kappa_{b}=2\pi\times5.25$ kHz, $\kappa_{c}=2\pi\times5.25$ kHz, and the average thermal photon numbers (in a dilution refrigerator around $15$ mK) are $\bar{n}_{a}\approx0$, $\bar{n}_{b}\approx0$, and $\bar{n}_{c}\approx0$.
Based on the above analysis, we know that the Fredkin-type interaction strength $g$ could reach $2\pi\times 1.3\times10^{3}$ kHz.
By controlling the driving fields, proper values of the driving detunings and magnitudes can be taken. Concretely, we choose $\Delta_{c}=2\pi\times 10$ MHz, which is a typical mechanical resonance frequency. We also choose  $\Delta_{b}/\Delta_{c}=2$ - $10$ to satisfy the approximation condition. The coupling amplification factors are taken as $|\xi_{b,c}|=50$ - $200$, which confirms that the coupling strengths ($g_{b,c}/\Delta_{c}=0.5$ - $2$) can enter the single-photon strong-coupling and even ultrastrong-coupling regimes. For the decay rates, we changed their values from experimental parameters to those larger than experimental parameters to show the influence of dissipation on the state generation in a wider parameter space. The fidelity of the state generation will be higher for lower decay rates. In Table~\ref{table1} we present the suggested parameters which are used in our simulations. By comparing these suggested parameters and the reported experimental parameters, we can expect that the physical implementation of our scheme should be within the reach of current or near future experimental conditions.

\section{Conclusion \label{Conclusion}}

We have proposed a scheme to simulate the three-mode optomechanical model, which consists of two mechanical-like modes and a single-mode optical field based on the Fredkin-type interaction. As an application of the simulated three-mode optomechanical interaction, we have studied the generation of entangled cat states in the two mechanical-like modes. The quantum properties of the generated states have been checked by calculating the joint Wigner function and the quantum entanglement. The influence of the dissipations on the state generation has also been analyzed with the quantum master equation method. In addition to the generation of macroscopic states, in this multimode optomechanical system, many interesting physical effects are still worth studying. We believe that our work will open up a different route to the study of few-photon optomechanical effects in multimode optomechanical systems.

\begin{acknowledgments}
J.-Q.L. was supported in part by National Science Foundation of China (Grants No.~11774087, No.~11822501, No.~12175061, and No.~11935006), Hunan Science and Technology Plan Project (Grant No.~2017XK2018), Innovation Training Project for College Students (Grant No. 2019055), and the Science and Technology Innovation Program of Human Province (Grants No.~2020RC4047 and No.~2021RC4029). J.-F.H. was supported in part by the National Natural Science Foundation of China (Grant No. 12075083), Scientific Research Fund of Hunan Provincial Education Department (Grant No. 18A007), and Natural Science Foundation of Hunan Province, China (Grant No. 2020JJ5345).
\end{acknowledgments}

\end{document}